\documentclass[prb,aps,twocolumn]{revtex4}
\usepackage{graphicx,amsfonts,bm,amsmath}
\newcommand{\bk}{b_{\bm{k}}}
\newcommand{\bkd}{b_{\bm{k}}^{\dag}}
\newcommand{\bmk}{b_{-\bm{k}}}
\newcommand{\bek}{\beta_{\bm{k}}}
\newcommand{\bemk}{\beta_{-\bm{k}}}
\newcommand{\bekd}{\beta_{\bm{k}}^{\dag}}
\newcommand{\sumk}{\sum_{\bm{k}}}
\newcommand{\wk}{\omega_{\bm{k}}}
\newcommand{\fk}{f_{n}(\bm{k})}

\newcommand{\kk}{\bm{k}}

\newcommand{\rl}{\rangle\!\langle}

\newcommand{\ff}{{\cal F}}
\DeclareMathOperator{\tr}{Tr}

\DeclareMathOperator{\princ}{\mathcal{P}\!\!}

\begin{document}
\title{Partly noiseless encoding of quantum information in quantum dot
arrays against phonon-induced pure dephasing}
\author{A. Grodecka}
 \email{Anna.Grodecka@pwr.wroc.pl}
\author{P. Machnikowski}
 \affiliation{Institute of Physics, Wroc{\l}aw University of
Technology, 50-370 Wroc{\l}aw, Poland}

\begin{abstract}
We show that pure dephasing of a quantum dot charge (excitonic) qubit may be
reduced for sufficiently slow gating by collectively encoding quantum
information in an array of quantum dots. We study the role of the size
and structure of the array and of the exciton
lifetime for the resulting total error of a single-qubit operation.
\end{abstract}

\pacs{}

\maketitle

\section{Introduction}

The idea of implementing quantum information
processing schemes on charge states in quantum dot (QD) systems
 \cite{zanardi98b,biolatti00} has recently been supported by an experimental
demonstration \cite{li03}. On the other hand, it has been proposed
that quantum bits encoded in electron spins (which have much longer
dephasing times) may be optically controlled by inducing charge
dynamics within various schemes exploiting selection rules and Pauli
blocking \cite{pazy03a,calarco03,chen04,troiani03}. Moreover, the
demonstration of a QD-based photodiode operating in
the quantum coherent regime \cite{zrenner02} shows great promise for
quantum optoelectronic applications. These experimental achievements
and theoretical proposals have driven a lot of investigation on
coherence properties of QD systems, in particular under external
driving of carrier dynamics. 

Limitations to the coherent control of confined carriers in QDs 
result, on one
side, from finite lifetime of charge states used for information
encoding; one restriction is imposed, for instance, by radiative
decay (on nanosecond time scale \cite{langbein04}) 
of confined excitons that seemed to be the most natural charge qubit
systems due to relative ease of ultrafast optical driving
\cite{biolatti00}. On the other side, optical experiments \cite{borri01} have
demonstrated that pure dephasing processes destroy coherence of
the system state to a large extent within a few picoseconds after
creating the state. This has been attributed to
carrier-phonon interaction \cite{krummheuer02,vagov03} and interpreted
in terms of spontaneous relaxation of the lattice after an optically induced
change of the confined charge distribution, accompanied by emission of
phonon wave packets \cite{vagov02a,jacak03b}, which creates a kind of
\textit{which way} trace in the crystal \cite{roszak06b} 
and results in dephasing. The same phonon response to the charge
evolution leads also to dephasing in optical spin control schemes
\cite{calarco03,roszak05b}. 

The phonon-induced dephasing effect may be avoided by
driving the system evolution adiabatically with respect to phonon
dynamics \cite{calarco03,alicki04a}, 
which requires long driving times and leads to 
growing probability of exciton decay during the operation. Optimal
driving conditions correspond to a tradeoff between these two
dephasing effects \cite{alicki04a}. As a
result, maximum fidelity of a coherent quantum control operation on a
given system is limited. 

One way to partly circumvent this limitation is optimizing the pulse shape
\cite{axt05a,hohenester04}. Another solution might be to modify the
properties of the phonon reservoir by placing the dot near the crystal
surface \cite{krummheuer05b} or in an acoustic cavity \cite{cazayous04}. 
An alternative approach is to overcome the physical restrictions
imposed by dephasing processes on the quantum-logical level. 
If the reservoir acts
collectively on a register composed of a number of charge qubits then 
subspaces exist which are safe against decoherence
\cite{zanardi97,bacon00,lidar01a,lidar01b}. In principle, 
this idea can be used to implement robust logical
qubits on QD arrays
\cite{zanardi98b,zanardi99a}. In practice, however, protection against 
decoherence based on collective interaction with the
environment (closely related to the subradiance effect) 
may be difficult to achieve in these artificial systems since it 
requires that the
transition energies in all QDs are the same. Moreover, the
phonon-induced pure
dephasing effect involves a range of modes selected by the system
geometry in such a way that reservoir correlation length roughly
corresponds to the single QD size which apparently precludes any
collective interaction with a number of QDs.

In this paper we show that, in spite of these difficulties, 
partly noiseless encoding against the pure dephasing effect is
possible. We focus our discussion on the case of an excitonic charge qubit.
In analogy to the original idea of noiseless
encoding\cite{zanardi97}, 
we will identify two states of an array of quantum dots which
span a subspace (logical qubit) that, under appropriate driving
conditions, allows performing single-qubit operations at reduced
error, i.e., with
relatively little lattice response, compared to switching of a single
excitonic physical qubit.
Physically, the collective encoding approach in this case relies on the
fact that the pure dephasing effect is of
dynamical character and results exclusively from nonadiabaticity of
the driving \cite{alicki04a}. For a given speed of rotation in
the qubit space, it involves 
only slow modes that cannot follow the evolution of
charge distribution, while faster modes follow
adiabatically and reversibly. In this way, only a long wave length part of all
the coupled modes is relevant for the dynamics. Since the range of
these slow modes is selected dynamically, independently of the system
geometry, it may correspond (under suitable driving conditions) to
wave lengths larger than the register (QD array) size, thus allowing for
collective interaction. 

In the present proposal, 
the logical qubit is
defined on a system of $N=2^{n}$ quantum dots (physical qubits) 
in such a way that the logical $|1\rangle$ state
corresponds to the presence of an exciton in each of the appropriately
chosen $N/2$ dots, while in the logical $|0\rangle$ state the other half of
the dots are occupied. 
With the correct definition of the logical
qubit, the excitations of the phonon reservoir induced by
the evolution of charges in all QDs interfere destructively in the
long wave length sector,
leading to a decrease of dephasing. 
The proposed scheme works only for relatively
slow evolution and the resulting fidelity gain is again restricted by
finite lifetimes. We will study how much the overall error may be
reduced depending on the exciton lifetime and the strength of other
decoherence processes. It
should be stressed that the scheme proposed here does not
assume that the individual dots in the register are identical. 

The paper is organized as follows. In Sec.~\ref{sec:model} we introduce
the model of the QD array implementing the logical qubit. Next, in
Sec.~\ref{sec:perturb}, we present the general method for describing
the effects of phonon perturbation on an arbitrary operation on a
logical qubit. Sec.~\ref{sec:collect} contains the results,
discussing the error reduction for various layouts of the logical
qubit. Sec.~\ref{sec:concl} concludes the paper with final
remarks. 

\section{Model}
\label{sec:model}

As a model of the physical QD array implementing the logical qubit,
we consider $N$ quantum dots, each approximated by a two-level
system corresponding to the absence or presence of an exciton
with a fixed polarization, with transition energies $\epsilon_{l}$,
$l=1,\ldots N$. The wave functions of carriers confined in
different dots do not overlap so that no phonon-assisted transfer is
possible. We will be interested in driving the system with long, i.e.,
spectrally narrow pulses so that excitations of optical phonons may be
excluded. If the electron and hole wave functions in each dot
overlap the piezoelectric coupling is very weak due to charge
cancellation \cite{krummheuer02}. 
Therefore, we assume that the confined carriers interact
only with longitudinal acoustic phonons via deformation potential coupling.
The Hamiltonian of this semiconductor system is then
\begin{eqnarray}\label{ham0}
H_{\mathrm{sc}} & =& 
\sum_{n}\epsilon_{n}a_{n}^{\dag}a_{n}+\sumk\hbar\wk\bkd\bk
\\ \nonumber 
&+&\sum_{n,\bm{k}}a_{n}^{\dag}a_{n}\fk\left(\bkd+\bmk \right),
\end{eqnarray}
where $a_{n}^{\dag},a_{n}$ are fermionic creation and annihilation
operators for the exciton in the $n$th dot, $\bkd,\bk$ are bosonic
operators for phonon modes, $\wk=ck$ are the corresponding frequencies
($c$ is the speed of sound),
and $\fk$ are the carrier-phonon coupling constants for an exciton in
the $n$th dot with the symmetry $\fk=f_{n}^{*}(\bm{-k})$. 
Assuming, for simplicity, that the exciton
state in the $n$th dot is described by 
a product of electron and hole wave
functions $\psi^{\mathrm{(e/h)}}_{n}(\bm{r}-\bm{r}_{n})$, 
where $\bm{r}_{n}$ is the
position of the dot,
the coupling constants may be written as \cite{grodecka05a}
\begin{equation}\label{fk}
\fk=\sqrt{\frac{\hbar k}{2\rho_{\mathrm{c}} v c}}
(\sigma_{\mathrm{e}}-\sigma_{\mathrm{h}})
e^{-i\bm{k}\cdot\bm{r}_{n}}{\cal F}_{n}(\bm{k}),
\end{equation}
with the formfactors
\begin{eqnarray}\label{ff}
{\cal F}_{n}(\bm{k}) & = & \frac{1}{\sigma_{\mathrm{e}}-\sigma_{\mathrm{h}}}
\int d^{3}rd^{3}r'
|\psi^{\mathrm{(e)}}_{n}(\bm{r})|^{2}
|\psi^{\mathrm{(h)}}_{n}(\bm{r}')|^{2}\\
\nonumber
&&\times\left(  \sigma_{\mathrm{e}}e^{-i\bm{k}\cdot\bm{r}} 
-\sigma_{\mathrm{h}}e^{-i\bm{k}\cdot\bm{r'}} \right).
\end{eqnarray}
Here $\rho_{\mathrm{c}}$ denotes the crystal density, 
$v$ is the normalization volume of the phonon modes,
and $\sigma_{\mathrm{e}/\mathrm{h}}$
are the deformation potential constants for electrons and holes.

The Hamiltonian $H_{\mathrm{sc}}$ [Eq.~(\ref{ham0})] has the structure of the
independent boson model and may be diagonalized exactly
\cite{mahan00}. To this end, one defines the operator 
\begin{displaymath}
\mathbb{W}
=\exp\left[\sum_{n,\bm{k}}a_{n}^{\dag}a_{n}g_{n}^{*}(\bm{k})\bk-\mathrm{H.c.}
\right],
\end{displaymath}
where $g_n(\bm{k})=\fk/(\hbar\wk)$,
and writes the Hamiltonian $H_{0}$ in terms of the operators 
\begin{equation}\label{alfa}
\alpha_{n}=\mathbb{W}^{\dag}a_{n}\mathbb{W}
=a_{n}W_{n}
\end{equation}
and
\begin{equation}\label{beta}
\bek=\mathbb{W}^{\dag}\bk\mathbb{W}
=\bk+\sum_{n}\alpha_{n}^{\dag}\alpha_{n}g_{n}(\bm{k}),
\end{equation} 
where
\begin{displaymath}
W_{n}=\exp\left[\sum_{\bm{k}}g^{*}_{n}(\bm{k})\bk-\mathrm{H.c.}\right].
\end{displaymath}
In the calculations one uses the fact that $[W_{n},W_{m}]=0$ for
non-overlapping exciton states.
The result is
\begin{equation}\label{ham-pol}
H_{\mathrm{sc}}=\sum_{n}E_{n}\alpha_{n}^{\dag}\alpha_{n}
+\sumk\hbar\wk\bekd\bek,
\end{equation}
where $E_{n}=\epsilon_{n}-\sumk|\fk|^{2}/(\hbar\wk)$. 
The new operators (\ref{alfa}) and (\ref{beta}) 
represent excitons in the dots surrounded by
coherent lattice displacement field and phonon modes shifted by
the presence of a charge distribution. 

We will associate the value $|0\rangle$ of the logical qubit with the state
of the QD
array in which a certain subset of $N/2$ dots, labelled by indices
$n_{1},\ldots n_{N/2}$, is occupied by
excitons and the value $|1\rangle$ with the state in which the
other $N/2$ ($n'_{1},\ldots n'_{N/2}$, with $n_{l}\neq n'_{l'}$ 
for all $l,l'=1,\ldots,N/2$) dots are occupied. 
Thus, the basis states of the logical qubit are
\begin{displaymath}
|0\rangle  =  a_{n_{1}}^{\dag}\ldots 
a_{n_{N/2}}^{\dag}|\mathrm{g}\rangle,\;\;\;\;
|1\rangle  =  a_{n'_{1}}^{\dag}\ldots 
a_{n'_{N/2}}^{\dag}|\mathrm{g}\rangle,
\end{displaymath}
where $|\mathrm{g}\rangle$ is the ground state of the system (with all
dots empty).
In our discussion we will assume that a control field is available
that couples the relevant pair of states of the multi-QD register and induces
a transition without leaving the subspace spanned by these states. 
In particular, we assume the availability of a control Hamiltonian of
the form 
\begin{equation}\label{ham-C}
H_{\mathrm{C}}=\frac{1}{2}f(t)e^{-i\Delta E t/\hbar}
a_{n_{1}}^{\dag}\ldots a_{n_{N/2}}^{\dag}a_{n'_{1}}\ldots a_{n'_{N/2}}
+\mathrm{H.c.},
\end{equation}
which transfers $N/2$ excitons from one subset of dots to
the other. Here $f(t)$ is the envelope of the control field and 
\begin{displaymath}
\Delta E=\sum_{l=1}^{N/2}E_{n_{l}}-\sum_{l=1}^{N/2}E_{n'_{l}}
\end{displaymath}
is the difference between the total energies of the logical states
$|0\rangle$ and $|1\rangle$.
A practical implementation of such a coupling is a separate
issue. For transferring a single
exciton between two dots, a realization of an effective
Hamiltonian of this kind has been proposed,
using the F{\"o}rster coupling and the
optical Stark effect \cite{lovett03a,lovett03b}. 

In terms of the redefined degrees of freedom [Eqs.~(\ref{alfa}) and
(\ref{beta})],
the control Hamiltonian (\ref{ham-C}) to the leading order in phonon
coupling reads
\begin{eqnarray}\label{ham-C-pol}
H_{\mathrm{C}} &= & \frac{1}{2}f(t)e^{-i\Delta E t/\hbar}
\alpha_{n_{1}}^{\dag}\ldots \alpha_{n_{N/2}}^{\dag}
\alpha_{n'_{1}}\ldots \alpha_{n'_{N/2}} \\
\nonumber
&& \times
\left[1+ \sumk G(\bm{k})(\bemk-\bekd)  \right] +\mathrm{H.c.}
\end{eqnarray}
with
\begin{displaymath}
G(\bm{k})=[g_{n'_{1}}(\bm{k})+\ldots +g_{n'_{N/2}}(\bm{k})]
-[g_{n_{1}}(\bm{k})+\ldots +g_{n_{N/2}}(\bm{k})].
\end{displaymath}
The phonon perturbation in Eq.~(\ref{ham-C-pol}) leads to reservoir
excitations accompanying any operation on the carrier subsystem but 
does not drive this subsystem outside the subspace spanned by the
logical qubit states. 

In our calculations the wave functions of confined electrons and holes
will be modelled by identical Gaussians,
\begin{displaymath}
\psi(\bm{r}) 
= \frac{1}{\pi^{3/4}l_z l}
\exp\left(-\frac{x^2+y^{2}}{2l^2}\right)
\exp\left(-\frac{z^2}{2l_z^2}\right),
\end{displaymath}
where the parameters $l$, $l_{z}$ are related to the QD size in the $xy$
plane and along the $z$ direction, respectively.
For simplicity, we assume here that the wave functions in all QDs are
identical. As shown in the Appendix, including a small variation of
dot sizes and relative electron and hole confinement widths 
will only lead to inessential quantitative corrections.
For this choice of wave functions, the formfactor [Eq.~(\ref{ff})]
has the explicit form
\begin{equation}\label{ff-av}
{\cal F}(\bm{k}) =
e^{-(k_{\bot} l/2)^2-(k_{z} l_{z}/2)^2 },
\end{equation}
where $k_{\bot}=(k_{x}^{2}+k_{y}^{2})^{1/2}$. 

We will assume that the operation is performed by a Gaussian control pulse,
\begin{equation}\label{gauss}
f(t)=\frac{\hbar\alpha}{\sqrt{2\pi}\tau_{\mathrm{p}}}
e^{-\frac{1}{2}(\frac{t}{\tau_\mathrm{p}})^2},
\end{equation}
where $\alpha$ is the angle determining the gate (corresponding to the
angle of rotation on the Bloch sphere) and $\tau_\mathrm{p}$
is the gate duration.

In Tab.~\ref{tab:param} the material parameters (corresponding to an
InAs/GaAs system) are collected.

\begin{table}
\begin{tabular}{lll}
\hline
Deformation potential coupling 
	& $\sigma_{\mathrm{e}}-\sigma_{\mathrm{h}}$ & 8 eV \\
Crystal density & $\rho_{\mathrm{c}}$ & 5360 kg/m$^3$ \\
Speed of sound (longitudinal) & $c$ & 5150 m/s \\
Long-time decoherence times\cite{langbein04} & $\tau_{0}$ & \\
at $T=0$ K & & 2 ns \\
at $T=10$ K & & 0.5 ns \\
\hline
\end{tabular}
\caption{\label{tab:param}System parameters used in the calculations.}
\end{table}

\section{Phonon perturbation during a quantum gate}
\label{sec:perturb}

In the interaction picture with respect to $H_{\mathrm{sc}}$
the system evolution is generated by the Hamiltonian
\begin{equation}\label{ham-rot}
H=\frac{1}{2}f(t)\sigma_{x}+\frac{1}{2}f(t)\sigma_{y}\hat{R}(t),
\end{equation}
where
\begin{displaymath}
\hat{R}(t)=i\sumk G(\bm{k})(e^{-i\wk t}\bemk-e^{i\wk t}\bekd)
\end{displaymath}
and $\sigma_{x,y}$ are Pauli matrices.

The evolution in the absence of phonon perturbation
is described by the unitary operator 
$U_{0}(t)$ generated by the first term in Eq.~(\ref{ham-rot}),
\begin{displaymath}
U_{0}(t)=\cos \frac{\Phi(t)}{2} 
-i\sin\frac{\Phi(t)}{2}\sigma_{x},\;\;\;
\Phi(t)=\frac{1}{\hbar}\int_{t_{0}}^{t}d\tau f(\tau).
\end{displaymath}

The effect of the phonon coupling on the driven dynamics may be
calculated using the second-order (Born) expansion of the evolution
equation for the density matrix \cite{alicki02a,roszak05b}, including
the first term in Eq.~(\ref{ham-rot}) exactly and the second one as
a perturbation.
The reduced density matrix of the qubit is written as 
\begin{equation}\label{dm-gener}
\rho(t)=U_{0}(t)[\rho_{0}+\rho^{(2)}(t)]U_{0}^{\dag}(t),
\end{equation}
where the correction $\rho^{(2)}(t)$ is calculated from a
perturbation expansion
\begin{eqnarray}\label{ro2}
    \lefteqn{\rho^{(2)}(t)=} \\
\nonumber & &
    -\frac{1}{\hbar^{2}}
	\int_{t_{0}}^{t}d\tau\int_{t_{0}}^{t}d\tau' \Theta(\tau-\tau')
      \tr_{\mathrm{R}}[\tilde{V}(\tau),[\tilde{V}(\tau'),\varrho_{0}]].
\end{eqnarray}
Here $\tr_{\mathrm{R}}$ is the trace with respect to the reservoir
(phonon) degrees of freedom and $\tilde{V}(t)=U_{0}^{\dag}(t)VU_{0}(t)$,
where $V$ is the second term in Eq.~(\ref{ham-rot}).
We will assume the initial state of the compound system in the form
$\varrho_{0}=\rho_{0}\otimes\rho_{T}$, where $\rho_{T}$ is the thermal 
equilibrium state of the phonon bath and the initial state of the qubit
subsystem is pure, $\rho_{0}=|\psi_{0}\rl\psi_{0}|$. 

Defining the operator
\begin{eqnarray*}
Y(\omega)=\frac{1}{\hbar}\int_{t_{0}}^{t}d\tau 
U_{0}^{\dag}(\tau)\frac{1}{2}f(\tau)\sigma_{y}U_{0}(\tau)e^{i\omega\tau}
\end{eqnarray*}
and the phonon spectral function
\begin{equation}\label{spdens}
R(\omega)= \frac{1}{2\pi}\int dt
  \langle \hat{R}(t)\hat{R}\rangle e^{i\omega t},
\end{equation}
and representing the Heaviside function as
\begin{displaymath}
\Theta(t)=-e^{i\omega t}\int\frac{d\omega'}{2\pi i}
   \frac{e^{-i\omega' t}}{\omega'-\omega+i0^{+}},
\end{displaymath}
one may write 
\begin{equation}\label{master}
\rho^{(2)}(t)= -i\left[ h_{t},\rho_{0} \right]
-\frac{1}{2}\left\{A_{t},\rho_{0}\right\}
+\hat{\Phi}_{t}[\rho_{0}],
\end{equation}
with 
\begin{eqnarray}\label{A}
    A_{t}& = & \int d\omega
    R(\omega)Y^{\dag}(\omega)Y(\omega),\\
\label{Fi}
    \hat{\Phi}_{t}\left[\rho\right] & = & 
    \int d\omega R(\omega) Y(\omega)\rho Y^{\dag}(\omega),\\
\label{h}
   h_{t}& = & \int d\omega R(\omega)
    \princ \int\frac{d\omega'}{2\pi}
    \frac{Y^{\dag}(\omega')Y(\omega')}{\omega'-\omega},
\end{eqnarray}
where $\mathcal{P}$ denotes the Cauchy principal value.

The explicit form of the operator $Y(\omega)$ is
\begin{displaymath}
Y(\omega) = \frac{i}{2}F(\omega)|+\rl -|
+\frac{i}{2}F^{*}(-\omega)|-\rl +|,
\end{displaymath}
where 
$|\pm\rangle=(|0\rangle\pm|1\rangle)/\sqrt{2}$ and
$F(\omega)$ is a nonlinear spectral characteristics of the control field,
\begin{equation}\label{F}
F(\omega)=\frac{1}{\hbar}\int_{-\infty}^{\infty}dt e^{i\omega t}f(t)e^{i\Phi(t)}.
\end{equation}
We assumed here that the initial time $t_{0}$ is before the control
field has been switched on and the final time $t$ is after it has been
switched off so that the limits of integration in the above formulas could be
extended to $\pm\infty$. 

To measure the effect of the phonon-induced dephasing on a
quantum gate we will use the \textit{error} 
$\delta$, defined as $\delta=1-F^{2}$, 
where 
$F=\langle\psi_{0}|U_{0}^{\dag}(\infty)
\rho(\infty) U_{0}(\infty)|\psi_{0}\rangle^{1/2}$ 
is the fidelity distance measure \cite{nielsen00} 
between the ideal (unperturbed) final state 
$U_{0}(\infty)|\psi_{0}\rl\psi_{0}|U_{0}^{\dag}(\infty)$ 
and the actual state $\rho(\infty)$. Substituting
Eq.~(\ref{dm-gener}) into this definition one finds
\begin{displaymath}
\delta=-\langle\psi_{0}|\rho^{(2)}(\infty)|\psi_{0}\rangle
=\int d\omega R(\omega)
|\langle \psi_{0}^{\bot}|Y(\omega)|\psi_{0}\rangle|^{2},
\end{displaymath}
where we used the explicit expression (\ref{master}) and 
$|\psi_{0}^{\bot}\rangle$ is a state orthogonal to 
$|\psi_{0}\rangle$ in the logical qubit space. Choosing
\begin{eqnarray*}
\psi_{0} & = & \sin\frac{\theta}{2}|+\rangle
  -e^{i\varphi}\cos\frac{\theta}{2}|-\rangle, \\
\psi^{\perp}_{0} & = & \cos\frac{\theta}{2}|+\rangle
  +e^{i\varphi}\sin\frac{\theta}{2}|-\rangle,
\end{eqnarray*}
one gets
\begin{eqnarray*}
\lefteqn{|\langle \psi_{0}^{\bot}|Y(\omega)|\psi_{0}\rangle|^{2}=}\\
&& \frac{1}{4}\left|F(\omega)e^{i\varphi}\cos^{2}\frac{\theta}{2}
-F^{*}(-\omega)e^{-i\varphi}\sin^{2}\frac{\theta}{2}\right|^{2}.
\end{eqnarray*}
Since, in general, the qubit may initially be in any state, we
average the error over the Bloch sphere of initial states, i.e., over
the angles $(\theta,\varphi)$, which yields
\begin{displaymath}
S(\omega)=
|\langle \psi_{0}^{\bot}|Y(\omega)|\psi_{0}\rangle|_{\mathrm{av}}^{2}
=\frac{1}{12}(|F(\omega)|^2+|F(-\omega)|^2).
\end{displaymath}
Since this is even in $\omega$, the error can be written in
the form
\begin{equation}\label{delta}
\delta=\int_{0}^{\infty}d\omega
\coth\left( \frac{\hbar\omega}{2k_{\mathrm{B}}T} \right) 
\frac{J(\omega)}{\omega^{2}}S(\omega),
\end{equation}
where 
\begin{equation}\label{J}
J(\omega)=\sumk|G(\bm{k})|^{2}\wk^2\delta(\omega-\wk)
\end{equation}
is the standard spectral density of the phonon reservoir.

\begin{figure}[tb] 
\begin{center} 
\unitlength 1mm
\begin{picture}(85,30)(0,6)
\put(0,0){\resizebox{85mm}{!}{\includegraphics{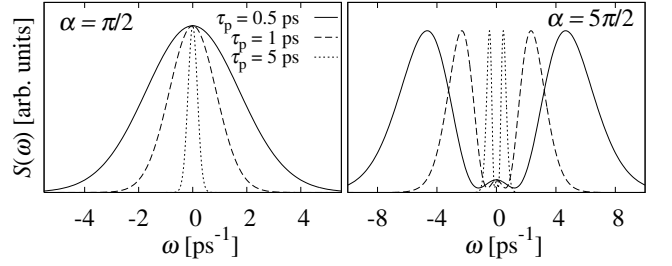}}}
\end{picture} 
\end{center} 
\caption{\label{fig:driving}
The spectral characteristics of the driving, $S(\omega)$, for two
rotation angles on the qubit Bloch sphere and for a series of pulse
durations as shown, for Gaussian pulses [Eq.~(\ref{gauss})]. The
scales on the two figures are different.}
\end{figure}

From Eq.~(\ref{F}) it can be seen that $F(\omega)$, and therefore also
$S(\omega)$, show simple scaling behavior with respect to the duration
of the control pulse. 
Namely, for a family of pulses of a fixed shape and area,
$f(t)=(\hbar\alpha/\tau_{0})\tilde{f}(t/\tau_{0})$, where $\tilde{f}(x)$ is
the fixed shape of the pulse and $\tau_{0}$ is the pulse duration, the
spectral function scales as
$S(\omega)=\tilde{S}(\tau_{0}\omega)$ (see Fig.~\ref{fig:driving}). 
Therefore, for
sufficiently long control pulses the error is determined by the low
frequency part of the phonon spectral density. To be specific, 
it can be seen from Eq.~(\ref{delta}) that if, for low frequencies,
$J(\omega)\sim\omega^{n}$ 
then $\delta\sim\tau_{0}^{-n+1}$ or $\delta\sim\tau_{0}^{-n+2}$ for
long enough pulses 
at low and high temperatures, respectively
\cite{alicki04a}. Obviously, slowing down the driving in order to
reduce the error is possible only to a certain extent, since for
longer durations the error becomes dominated by decoherence effects related,
e.g., to finite exciton lifetime and phonon-assisted transitions,
which accumulate with time (linearly for short times). The interplay
of these two sources of dephasing creates a trade-off situation and
sets a \textit{lower bound} on the total error achievable for various
pulse durations \cite{alicki04a}. 

Although, in general, the phonon spectral
density for a single exciton confined in a QD depends on the dot
geometry, the low frequency behavior is only material-specific (see
Appendix) and can only be modified by engineering the properties of
phonon modes \cite{krummheuer05b,cazayous04}. 
However, as we show in the next Section, the
low-frequency behavior of the effective phonon spectral density may 
be modified
by collective encoding in an array of QDs, which allows one to reduce
the total error.

\section{Error reduction by collective encoding}
\label{sec:collect}

\begin{figure}[tb] 
\begin{center} 
\unitlength 1mm
\begin{picture}(85,40)(0,6)
\put(0,0){\resizebox{85mm}{!}{\includegraphics{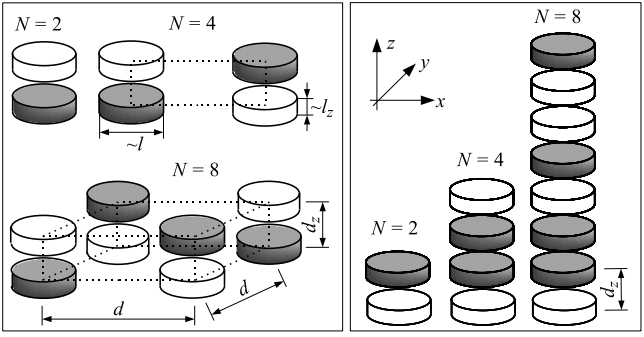}}}
\end{picture} 
\end{center} 
\caption{\label{fig:layout}
The two layouts of the QD array: 3-dimensional (left) and linear
(right). The dark dots are occupied in the logical $|1\rangle$ state
and the white dots in the $|0\rangle$ state. In the linear layout, the
logical qubit with $2N$ dots is constructed by interchanging occupied
and empty dots in the $N$ dot qubit and appending it to the original one.}
\end{figure}

We consider two kinds of arrays of quantum dots. The first one,
referred to as \textit{3-dimensional} (or \textit{3D}), is formed by
stacking 2 QDs along the strong confinement ($z$) axis, arranging 4
QDs in a rectangle in the $xz$ plane or placing 8 QDs in the corners
of a cuboid (Fig.~\ref{fig:layout}, left). In the second layout, which we
will refer to as \textit{linear}, $N=2^{n}$ dots are arranged along
the $z$ axis, as shown and explained in the right part of
Fig.~\ref{fig:layout}. 

\begin{figure}[tb] 
\begin{center} 
\unitlength 1mm
\begin{picture}(85,55)(0,6)
\put(0,0){\resizebox{85mm}{!}{\includegraphics{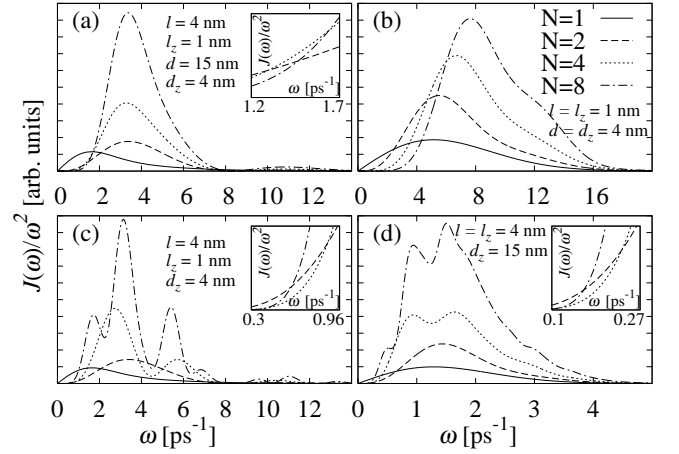}}}
\end{picture} 
\end{center} 
\caption{\label{fig:spectral}
The phonon spectral densities for the physical excitonic qubit (solid
line) and for logical qubits composed of $N$ physical qubits (dashed
lines) for two kinds of QD arrays: 3-dimensional (a,b) and linear
(c,d) with geometrical parameters as indicated.}
\end{figure}

For $N=2^{n}$ quantum dots in the 3D layout one finds the coupling
constants $G_{n}$ (for an appropriate choice of origin)
\begin{eqnarray*}
G_{1}(\kk) & = & 2i\sin\frac{k_{z}d_{z}}{2}G_{0}(\kk),\\
G_{2}(\kk) & = & 
(2i)^{2}\sin\frac{k_{z}d_{z}}{2}\sin\frac{k_{x}d}{2}G_{0}(\kk),\\
G_{3}(\kk) & = & 
(2i)^{3}\sin\frac{k_{z}d_{z}}{2}\sin\frac{k_{x}d}{2}
\sin\frac{k_{y}d}{2}G_{0}(\kk),
\end{eqnarray*}
where 
\begin{displaymath}
G_{0}(\kk)=\sqrt{\frac{1}{2\hbar\rho_{\mathrm{c}}vc^{3}k}}
(\sigma_{\mathrm{e}}-\sigma_{\mathrm{h}}){\cal F}(\kk)
\end{displaymath}
is the coupling constant for a single dot placed at the origin and the
distance parameters $d,d_{z}$ are defined in Fig.~\ref{fig:layout}.
From Eq.~(\ref{J}) one then finds the phonon spectral densities for the
logical qubits in the low frequency sector,
\begin{subequations}
\begin{eqnarray}\label{J-3d}
J_{1}(\omega) & \approx & J_{0}(\omega)\frac{d_{z}^{2}}{3c^{2}}\omega^{2} 
\approx \frac{\hbar (\sigma_{\mathrm{e}}-\sigma_{\mathrm{h}})^2 d_{z}^{2}}{12 \pi^{2} \rho_{\mathrm{c}}
c^{7}}\omega^{5}, \\
J_{2}(\omega) & \approx & J_{0}(\omega)\frac{d_{z}^{2}
d^{2}}{15c^{4}}\omega^{4} 
 \approx  \frac{\hbar(\sigma_{\mathrm{e}}-\sigma_{\mathrm{h}})^2 d_{z}^{2} d^{2}}
{60 \pi^{2} \rho_{\mathrm{c}}c^{9}}\omega^{7}, \\
J_{3}(\omega) & \approx & J_{0}(\omega)\frac{d_{z}^{2}
d^{4}}{105c^{6}}\omega^{6} 
 \approx  \frac{\hbar(\sigma_{\mathrm{e}}-\sigma_{\mathrm{h}})^2 d_{z}^{2} d^{4}}
{420 \pi^{2} \rho_{\mathrm{c}}c^{11}}\omega^{9},
\end{eqnarray}
\end{subequations}
where 
\begin{displaymath}
J_{0}(\omega) \approx \frac{\hbar (\sigma_{\mathrm{e}}-\sigma_{\mathrm{h}})^2}{4 \pi^2 \rho_{\mathrm{c}} c^5}\omega^3.
\end{displaymath}
These spectral densities are plotted in Fig.~\ref{fig:spectral}(a,b). 

For the linear array of QDs one finds
\begin{equation}
G_n(\bm{k}) = (2i)^n \prod_{j=-1}^{n-2} \sin{(2^j k_z d_{z})}G_{0}(\kk),
\end{equation}
which leads to the spectral density $J_{1}(\omega)$ as above and
\begin{subequations}
\begin{eqnarray}\label{J-lin}
J_{2}(\omega) & \approx & J_{0}(\omega)\frac{4 d_{z}^{4}}{5 c^{4}}\omega^{4} \approx \frac{\hbar(\sigma_{\mathrm{e}}-\sigma_{\mathrm{h}})^2 d_{z}^{4}}
{5 \pi^{2} \rho_{\mathrm{c}}c^{9}}\omega^{7}, \\
J_{3}(\omega) & \approx & J_{0}(\omega)\frac{64d_{z}^{6}}{7c^{6}}\omega^{6} \approx \frac{16 \hbar(\sigma_{\mathrm{e}}-\sigma_{\mathrm{h}})^2 d_{z}^{6}}
{7 \pi^{2} \rho_{\mathrm{c}}c^{11}}\omega^{9},
\end{eqnarray}
\end{subequations}
shown in Fig.~\ref{fig:spectral}(c,d).

In both layouts, $J_{n}(\omega)\sim\omega^{2n+3}$ for small $\omega$ 
so that for larger logical qubits the spectral density vanishes more
quickly for $\omega\to 0$, 
which will result in a faster decrease of the pure dephasing
error with growing duration of the control pulse, as discussed in the
previous Section. In order to see this, we
calculate the phonon-induced error using Eq.~(\ref{delta}) with the
appropriate spectral density $J_{n}(\omega)$ [obtained from
Eq.~(\ref{J}), for arbitrary $\omega$] for various logical
qubits. 

\begin{figure}[tb] 
\begin{center} 
\unitlength 1mm
\begin{picture}(85,55)(0,6)
\put(0,0){\resizebox{85mm}{!}{\includegraphics{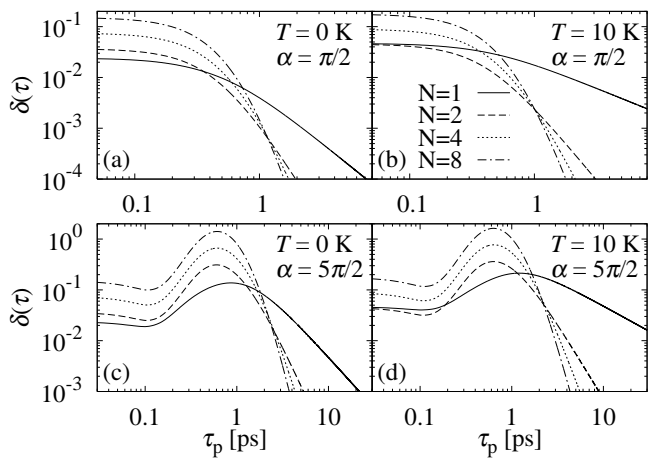}}}
\end{picture} 
\end{center} 
\caption{\label{fig:dyn3D}
The phonon-induced dynamical error for a $\pi/2$ and $5\pi/2$ rotation
on a physical qubit (solid line) and on a 3D logical qubit
with 2,4, and 8 dots (dashed and dotted lines) 
with sizes $l=4$ nm, $l_{z}=1$ nm and separations
$d=15$ nm, $d_{z}=4$ nm, at two different temperatures as shown.}
\end{figure}

Let us start the discussion with a logical qubit in the 3D layout,
composed of QDs of dimensions typical for self-assembled systems
(Fig.~\ref{fig:dyn3D}). First, let us consider the $\pi/2$ rotation of
the qubit (i.e., the single-qubit gate $e^{-i\pi\sigma_{y}/4}$).
For ultrafast pulses
($\tau_{\mathrm{p}}\lesssim 100$ fs), in most cases 
the error grows with the number
of QDs in the array. This regime corresponds to very broad spectral
functions $S(\omega)$ (see Fig.~\ref{fig:driving}), 
so that the whole range of phonon modes
contribute to the dephasing. Since considerable destructive
interference of phonon excitations is
unlikely to appear over the whole wide frequency sector of the phonon
spectral density, it may be expected that driving a
number of dots simultaneously should lead to more excitation of the
phonon reservoir and to more dephasing. In terms of the spectral
densities (Fig.~\ref{fig:spectral}), this effect is reflected by the
growing overall magnitude of $J(\omega)$. 

This situation changes for longer pulses ($\tau_{\mathrm{p}}\sim 1$
ps). Now only the long wavelength phonons
contribute to dephasing. From the general discussion presented above
it is clear that for sufficiently long pulses large collective qubits
will lead to lower phonon-induced error than smaller ones due to more
favorable scaling of the error with the pulse duration. Indeed, in
Fig.~\ref{fig:dyn3D} the value of $\delta$ decreases for long enough
pulses ($\tau_{\mathrm{p}}\gtrsim 1$ ps) according to a power-law
dependence with higher negative exponents for larger arrays.

It should be stressed that the reduction of dephasing achieved with
collective encoding is not just due to extending the effective
system size and the resulting weakening of the overall phonon coupling
\cite{krummheuer02,krummheuer05}. Such an effect might quantitatively
reduce dephasing but could not change the exponent of the
low-frequency behavior of $J(\omega)$. Moreover, the reduction of
$J(\omega)$ at $\omega\to 0$ appears only for the special definitions
of logical qubits shown in Fig.~\ref{fig:layout}. For a
different assignment of physical QDs to the logical qubit states, the
dephasing may even be increased, which shows that the effect is
actually due to destructive interference between the reservoir
excitations caused by the simultaneous switching of the appropriately
arranged physical qubits.

For large angles $\alpha$, corresponding to full Rabi rotations around
the Bloch sphere, the error dependence on the pulse duration becomes
more complicated due to the presence of many maxima in $S(\omega)$
(Fig.~\ref{fig:driving}).
While the power-law decrease of $\delta$ at $\tau_{\mathrm{p}}\gtrsim 1$ ps
corresponds to the situation when all of $S(\omega)$ lies in the
low-frequency sector of $J(\omega)$, the minimum at
$\tau_{\mathrm{p}}\sim 100$ fs appears when the area of large
$J(\omega)$ coincides with the local minimum between the maxima of
$S(\omega)$. Since the collective encoding improves only the
low-frequency behavior of $J(\omega)$ only the former is
advantageously affected by increasing the size of the QD array.

\begin{figure}[tb] 
\begin{center} 
\unitlength 1mm
\begin{picture}(85,77)(0,6)
\put(0,0){\resizebox{85mm}{!}{\includegraphics{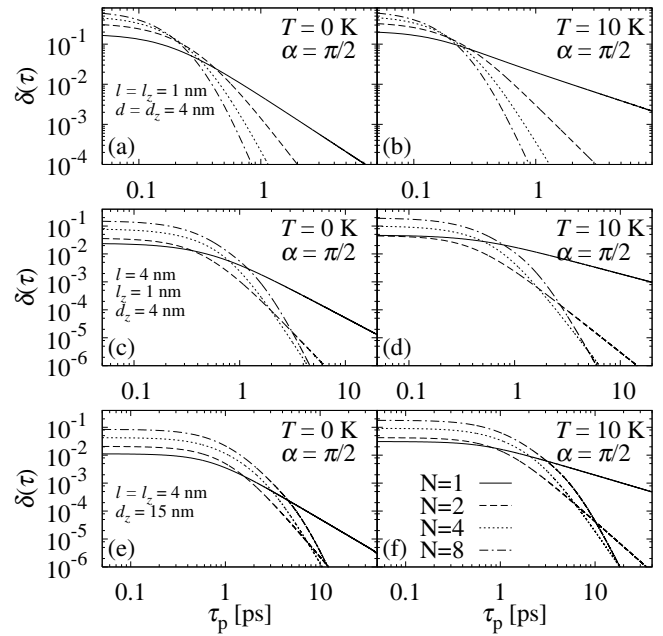}}}
\end{picture} 
\end{center} 
\caption{\label{fig:layouts}
Comparison of the dynamical error for three different geometries of
the QD array: small, closely spaced dots in the 3D layout (a,b), flat
dots of moderate size (like in Fig.~\ref{fig:dyn3D}) 
in the linear layout (c,d) and large dots in the
linear layout (e,f).}
\end{figure}

In the ultrafast limit, the dephasing of a single physical excitonic
qubit increases for decreasing dot sizes, since the $1/l$ momentum cut-off
in the formfactor [Eq.~(\ref{ff-av})] is shifted towards larger momenta
and the total area of the phonon spectral density grows. This can 
also be seen in Fig.~\ref{fig:spectral}(b),
corresponding to QDs with the lateral size reduced to $l=1$ nm. For
longer pulses, this difference in $J(\omega)$ becomes less important
since only the low-frequency part of this spectral density is
relevant. Indeed, comparing Fig.~\ref{fig:layouts}(a,b) 
(corresponding to smaller dots) with 
Fig.~\ref{fig:dyn3D}(a,b) one notes that the error for ultrashort pulses
is much larger for smaller dots but in the region of
$\tau_{\mathrm{p}}\gtrsim 1$ ps the error for a single
dot (solid line) grows very
little when the dot size is reduced. On the other hand, smaller dots
allow smaller distances $d$ between the physical qubits in the array. As a
result, the modulation of $J(\omega)$ resulting from the multi-QD
interference attains a longer period $\Delta\omega\sim c/d$ and the
sector of low frequencies, where the spectral densities $J_{n}(\omega)$
are ordered according to their asymptotic power-law behavior,
extends [see Fig.~\ref{fig:spectral}(b)]. Now, 
the improved behavior of the spectral
densities is fully reflected by the reduced values of errors for
growing number of QDs in the array already for 
$\tau_{\mathrm{p}}\gtrsim 300$ fs.

The linear layout allows the dots to be tightly stacked along the direction
of smallest size (growth direction of the semiconductor structure)
but, on the other hand, results in a growing distance
between the physical qubits on the opposite ends of the chain. This
leads to a quickly growing pre-factor at the low-frequency power-law
expression [compare Eqs.~(\ref{J-3d}) and (\ref{J-lin})]. 
As a result, the spectral densities for larger arrays
cross $J_{2}(\omega)$ at very low frequencies (see insets in
Fig.~\ref{fig:spectral}) and one needs 
longer pulses to take advantage of the lower values of $J_{n}(\omega)$
for $n>1$, as can be seen in Fig.~\ref{fig:spectral}(c,d).

Obviously, the quantitative results presented here depend on the
interplay of the physical coupling
parameters and the  wave number dependence of the original (physical)
coupling in the long wave length limit.
In any case, however, the spacing between the dots is crucial. This is
illustrated by Fig.~\ref{fig:layouts}(e,f), where the total error for a
linear array of large and widely spaced dots is plotted. From the
spectral density shown in Fig.~\ref{fig:spectral}(d) it is clear that
such a geometry leads not only to the overall narrowing of the phonon
spectral characteristics (due to large size) but also to the squeezing of
the region when the functions $J(\omega)$ show the advantageous
low-frequency behavior. As a result, increasing array size leads to
error reduction only for very long pulses.

In a real system, the dynamically induced pure dephasing is not the
only error source. Additional error comes from the finite exciton
lifetime and other processes, including phonon-assisted transitions to
excited exciton states \cite{borri01,borri02b}, dephasing by phonon
scattering \cite{muljarov04,machnikowski05c} and probably other 
effects whose exact nature seems to
be poorly understood. The common feature of all these dephasing
sources is that they can be described within a Markovian model leading
to an exponential decay of coherence with a single time constant
$\tau_{0}$ which is experimentally accessible via exciton line
broadening. For $\tau_{\mathrm{p}}\ll\tau_{0}$ their contribution to
the overall error grows linearly with the gate duration
$\tau_\mathrm{p}$. Therefore, we add a contribution of 
$\delta_{0}=\tau_{\mathrm{p}}/\tau_{0}$ to the total error. 
For brevity, this additional contribution will be referred to as related to
``finite lifetime''.
Depending on the physical process underlying this error, the time
constant $\tau_{0}$ may depend in various ways on the number of
physical qubits in the array. Here, we will consider two scenarios.
In the first one, $\tau_{0}$ is independent of $N$ (which assumes some
kind of collective action of the dephasing), while in the
second one it is proportional to $1/N$ (which corresponds to
individual decoherence of the physical qubits).

\begin{figure}[tb] 
\begin{center} 
\unitlength 1mm
\begin{picture}(85,55)(0,6)
\put(0,0){\resizebox{85mm}{!}{\includegraphics{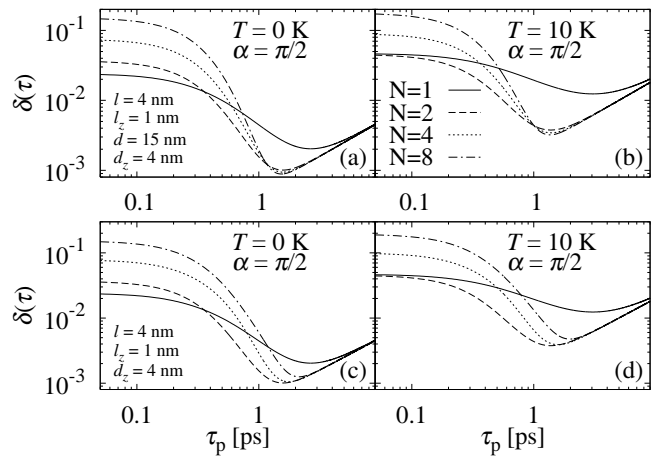}}}
\end{picture} 
\end{center} 
\caption{\label{fig:tau}
The total error, including a contribution from finite exciton lifetime
and other long-time scale decoherence processes for typical flat dots
(with sizes as shown) in 3D (a,b) and linear (c,d) layout at two
different temperatures.}
\end{figure}

Reaching the low frequency regime, necessary for reducing the
dynamical error, requires long control pulses and leads to 
growing values of $\delta_{0}$. As a result, for long pulse durations
the error is always dominated by the linearly growing finite lifetime
contribution. The interplay of the dynamical phonon-induced error 
(growing for short control pulses) and the finite lifetime decoherence
(increasing for long pulses) leads to well-defined driving conditions
where the error has its minimum \cite{alicki04a}. As shown in
Fig.~\ref{fig:tau}, in the first model of decoherence 
(with lifetimes $\tau_{0}=2$ ns and 0.5 ns at $T=0$ K and 10 K,
respectively \cite{langbein04}) the optimal total error is reduced by
a factor of 2 at $T=0$ K and almost by a factor of 4 
at $T=10$ K. It is interesting to note that for these life times large
logical qubits  
$N>2$ bring very little improvement over the simplest one ($N=2$) and
only in the 3D layout. 
  
\begin{figure}[tb] 
\begin{center} 
\unitlength 1mm
\begin{picture}(85,55)(0,6)
\put(0,0){\resizebox{85mm}{!}{\includegraphics{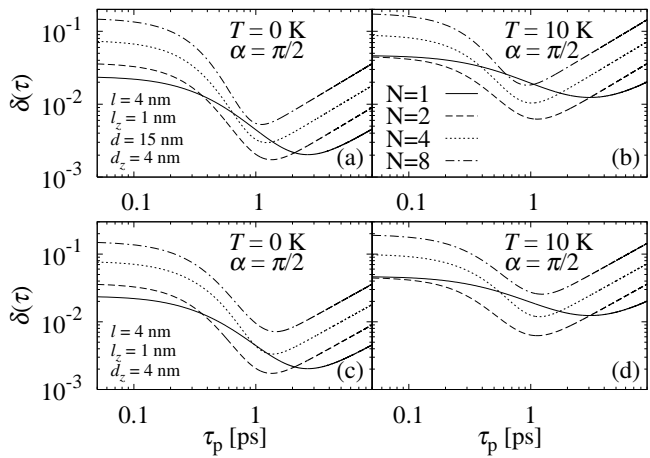}}}
\end{picture} 
\end{center} 
\caption{\label{fig:tau-n}
As in Fig.~\ref{fig:tau} but with decay probability proportional to the
number of physical qubits.}
\end{figure}

For the second decoherence model, the reduction of the effective
lifetime of the array considerably restricts the possible error reduction. At
$T=0$ one obtains only a fidelity gain of 17\%. More error reduction
(by a factor of 2) is achievable at higher temperatures. Due to large
increase of the error with growing number of physical qubits, only the
encoding in $N=2$ physical qubits brings the desired effect.

\begin{figure}[tbp] 
\begin{center} 
\unitlength 1mm
\begin{picture}(85,55)(0,6)
\put(0,0){\resizebox{85mm}{!}{\includegraphics{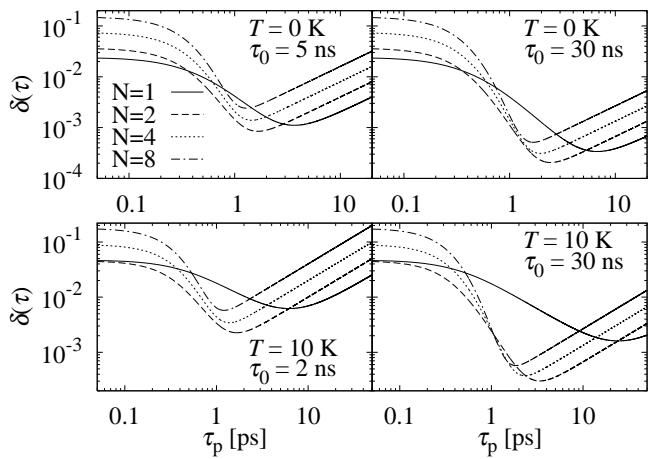}}}
\end{picture} 
\end{center} 
\caption{\label{fig:xyz-big-decoh}
The total error for a 3D logical qubit with geometry as in
Fig.~\ref{fig:tau-n}(a,b) for (hypothetical) long characteristic times of
the exciton lifetime, with decay-related error proportional to the
number of physical qubits, for $\alpha=\pi/2$ at two temperatures.}
\end{figure}

Thus, nanosecond decoherence times observed in the currently manufactured
structures strongly restrict the degree of reduction of dephasing
achievable with collective encoding, especially in the independent
decoherence model. 
For realistic sizes and
separations of the QDs in the array, taking advantage of the reduced
spectral density requires pulses of durations exceeding 1 ps, which
immediately leads to an error of order of $10^{-3}$, further growing for
increasing number of dots in the array. However, rapid progress of QD
engineering may lead to systems with considerably longer coherence
times, as suggested by the observation of long-lived excitons 
in particular QD systems \cite{lundstrom99}. 
An advantage of the collective qubit is that such reduction of
the long time decoherence is much more effectively reflected by the
reduction of the total error. As can be seen in
Fig.~\ref{fig:xyz-big-decoh}, extending the exciton lifetime leads to
a reduced error both for a single physical qubit and for a collective
qubit. However, the error reduction is much larger for the logical
qubit. For instance, the 2-QD encoding
(Fig. \ref{fig:tau-n}) at $T=0$ reduces the error by 17\% for $\tau_{0}=2$
ns, which grows to 31\% if the lifetime is extended to 5 ns and to a
factor of almost 1.7 for $\tau_{0}=30$ ns. There is even more gain at higher
temperatures. For $T=10$ K, one obtains an error reduction by a factor of
2.8 if the lifetime is extended to 2 ns (which is the actual
radiatively limited value in some structures\cite{langbein04}) and by
almost an order of magnitude for $\tau_{0}=30$ ns.

\section{Conclusions and outlook}
\label{sec:concl}

The discussion presented above shows that engineering of the effective
phonon spectral density by collective encoding of a logical qubit in an
array of physical (excitonic) qubits may lead to a reduction of the
phonon-induced dephasing during control operations on the qubit. This
is possible since for long enough pulses (slow driving) only
long-wavelength part of the phonon reservoir is involved in the
dephasing dynamics which extends the effective correlation length of
the reservoir and opens the possibility for collective encoding. Such
collective encoding reduces the total error resulting from the joint
action of various dephasing mechanisms and
increases the gain from possible further extension of the exciton life
times. With the collective encoding it seems feasible to reduce the
total error for a single-qubit operation to values of order of
$10^{-3}$ even at the relatively high temperature of $T=10$ K, which
is remarkable in view of the recent progress in the quantum error
correction theory under realistically strong dephasing
\cite{knill05}. 

The present analysis was restricted to Gaussian pulses. It is known,
however, that pulse shaping \cite{hohenester04} or pulsed control
techniques \cite{axt05a} lead to reduced phonon dephasing effect. The
reach structure of the effective phonon spectral density for
collective qubits suggests that 
combining the pulse shaping techniques with the
collective encoding may be very fruitful. Another possibility of reducing
the overall error might be to combine the collective encoding against
dynamical dephasing with the noiseless (subradiant) encoding against
radiative decay \cite{zanardi97}.

In more general terms, our discussion shows that certain details of
the phonon spectral density may be of essential importance for the
dephasing of a system driven by finite pulses. Obviously, the system
response to an ultrashort pulse, as observed in optical experiments
\cite{borri01},  yields valuable information on the
overall strength of the phonon-induced dephasing. However, we have
seen that the degree of decoherence for finite pulses is governed only
by the low-frequency part of the relevant spectral function and may
decrease even though the dephasing in the ultrafast limit
increases (e.g., for arrays of smaller but more closely spaced QDs). 
It is therefore possible that systems based on various
materials \cite{krummheuer05} and restricted geometries
\cite{krummheuer05b} may offer previously unnoticed possibilities
in the slow driving regime. Understanding of the peculiarities of the
carrier-phonon dynamics underlying the dephasing of driven
systems may also be helpful for designing robust schemes for the optical
control of spin qubits.

Supported by the Polish Ministry of Scientific Research
and Information Technology (PBZ-MIN-008/P03/2003). 
P.M. is grateful to A. von Humboldt Foundation for support.

\begin{appendix}
\section{Universality of carrier-phonon coupling in the long wave
length limit}

In this Appendix we show that for small variation of individual dot
sizes and for small difference between the electron and hole
localization in a dot, the exciton-phonon coupling constant 
can be written in the form (\ref{fk}), with the formfactor
replaced by Eq.~(\ref{ff-av}) which depends only on the average 
width of the electron and hole wave functions in the QD array.

In general, the electron and hole wave functions in the $n$th
dot may be characterized by different localization widths 
$l^{(\mathrm{e/h})}_{n}$ (for simplicity, we still assume that the
exciton wave function is a product of Gaussians
isotropic in the $xy$ plane and with a fixed width $l_{z}$). The
carrier-phonon interaction is then described by Eq.~(\ref{fk}) with
\cite{grodecka05a}
\begin{displaymath}
\ff_{n}(\kk) = 
\widetilde\sigma_{\mathrm{e}}
e^{-k_{\bot}^{2} l_{n}^{\mathrm{(e)}2}/4
-k_{z}^{2} l_{z}^{2}/4}
-\widetilde\sigma_{\mathrm{h}}
e^{-k_{\bot}^{2} l_{n}^{\mathrm{(h)}2}/4
-k_{z}^{2} l_{z}^{2}/4},
\end{displaymath}
where $\widetilde{\sigma}_{\mathrm{e/h}}
=\sigma_{\mathrm{e/h}}/(\sigma_{\mathrm{e}}-\sigma_{\mathrm{h}})$.
We define the averaged wave function width
\begin{displaymath}
l^{2}
=\frac{1}{2N}\left[
\sum_{n}(l_{n}^{\mathrm{(e)}})^{2}+\sum_{n}(l_{n}^{\mathrm{(h)}})^{2}
\right]
\end{displaymath}
and the deviations from the average
\begin{displaymath}
\left(\Delta l_{n}^{\mathrm{(e/h)}}\right)^{2}
=\left(l_{n}^{\mathrm{(e/h)}}\right)^{2}-l^{2}.
\end{displaymath}
The formfactor can now be written as 
\begin{eqnarray*}
\ff_{n}(\kk) & = & 
e^{-k_{\bot}^{2} l^{\mathrm{2}}/4 -k_{z}^{2} l_{z}^{2}/4}\\
&&\times\left[
\widetilde\sigma_{\mathrm{e}}
e^{-k_{\bot}^{2} (\Delta l_{n}^{\mathrm{(e)}})^{2}/4}
-\widetilde\sigma_{\mathrm{h}}
e^{-k_{\bot}^{2} (\Delta l_{n}^{\mathrm{(h)}})^{2}/4}
 \right].
\end{eqnarray*}
It is clear that $k_{\bot}$ is cut off at the value of $\sim 1/l$, hence
$k_{\bot}\Delta l_{n}^{\mathrm{(e/h)}} \ll 1$
if $\Delta l_{n}^{\mathrm{(e/h)}}\ll l$, i.e., when the variation of
dot sizes is relatively small. The exponents may be therefore expanded
into power series,
\begin{eqnarray*}
\ff_{n}(\kk) & = & 
e^{-k_{\bot}^{2} l^{\mathrm{2}}/4 -k_{z}^{2} l_{z}^{2}/4}\\
&&\times \left[1-
\widetilde\sigma_{\mathrm{e}}
\frac{k_{\bot}^{2} (\Delta l_{n}^{\mathrm{(e)}})^{2}}{4}
-\widetilde\sigma_{\mathrm{h}}
\frac{k_{\bot}^{2} (\Delta l_{n}^{\mathrm{(h)}})^{2}}{4}
 \right],
\end{eqnarray*}
where we took into account that 
$\widetilde{\sigma}_{\mathrm{e}}-\widetilde{\sigma}_{\mathrm{h}}=1$.
Retaining only the leading term one arrives at Eq.~(\ref{ff-av}).
It should be noted that independently of the shape of the wave
functions, normalization condition requires that for long wavelengths
$\ff_{n}(\kk)=1+O(k^{2})$, which is sufficient for deriving
Eqs.~(\ref{J-3d}-c) and (\ref{J-lin},b).

\end{appendix}


\end{document}